\documentclass[12pt]{iopart}
\usepackage{amsthm}
\usepackage{amsmath}
\usepackage{amssymb}
\usepackage{mathtools}
\usepackage{graphicx}
\usepackage[left=25mm,right=25mm,top=25mm,columnsep=15pt]{geometry} 
\usepackage[T1]{fontenc}
\usepackage[colorlinks = true, allcolors = blue]{hyperref}
\usepackage[subrefformat=parens,labelformat=parens,caption=false]{subfig}
\usepackage[square,sort&compress,comma,numbers]{natbib}
\begin{document}
\title[Pulsed multireservoir engineering for a trapped ion]{Pulsed multireservoir engineering for a trapped ion with applications to state synthesis and quantum Otto cycles}

\author{W S Teixeira$^1$\footnote{Present address:
QCD Labs, QTF Centre of Excellence, Department of Applied Physics, Aalto University, P.O. Box 15100, FI-00076 Aalto, Finland.}, M K Keller$^2$, 
F L Semi\~ao$^1$}

\address{$^1$ Centro de Ci\^encias Naturais e Humanas, Universidade Federal do ABC, Santo Andr\'e, 09210-170 S\~ao Paulo, Brazil}
\address{$^2$ Department of Physics and Astronomy, University of Sussex, Brighton BN1 9RH, United Kingdom}
\ead{fernando.semiao@ufabc.edu.br}

\begin{abstract}
Conducting an open quantum system towards a desired steady state through reservoir engineering is a remarkable task that takes dissipation and decoherence as tools rather than impediments. Here we develop a collisional model to implement reservoir engineering for the one-dimensional harmonic motion of a trapped ion. Our scheme is based on the pulsed interaction between the vibrational mode and the electronic levels of a trapped ion, which is promoted by resolved-sideband lasers. Having multiple internal levels, we show that multiple reservoirs can be engineered, allowing for more efficient synthesis of well-known non-classical states of motion and the generation of states that are unfeasible with a single-bath setup, for instance, thermal states with arbitrary positive temperatures. We apply these ideas to quantum Otto cycles beyond purely thermal reservoirs. In particular, we present general conditions for the violation of the standard Otto bound in the limiting regime of non-adiabatic dynamics. 
\end{abstract}


\global\long\def\ket#1{|#1\rangle}%

\global\long\def\Ket#1{\left|#1\right>}%

\global\long\def\bra#1{\langle#1|}%

\global\long\def\Bra#1{\left<#1\right|}%

\global\long\def\bk#1#2{\langle#1|#2\rangle}%

\global\long\def\BK#1#2{\left\langle #1\middle|#2\right\rangle }%

\global\long\def\kb#1#2{\ket{#1}\!\bra{#2}}%

\global\long\def\KB#1#2{\Ket{#1}\!\Bra{#2}}%

\global\long\def\mel#1#2#3{\bra{#1}#2\ket{#3}}%

\global\long\def\MEL#1#2#3{\Bra{#1}#2\Ket{#3}}%

\global\long\def\n#1{|#1|}%

\global\long\def\N#1{\left|#1\right|}%

\global\long\def\ns#1{|#1|^{2}}%

\global\long\def\NS#1{\left|#1\right|^{2}}%

\global\long\def\nn#1{\lVert#1\rVert}%

\global\long\def\NN#1{\left\lVert #1\right\rVert }%

\global\long\def\nns#1{\lVert#1\rVert^{2}}%

\global\long\def\NNS#1{\left\lVert #1\right\rVert ^{2}}%

\global\long\def\ev#1{\langle#1\rangle}%

\global\long\def\EV#1{\left\langle #1\right\rangle }%

\global\long\def\ha{\hat{a}}%

\global\long\def\hb{\hat{b}}%

\global\long\def\hc{\hat{c}}%

\global\long\def\hd{\hat{d}}%

\global\long\def\he{\hat{e}}%

\global\long\def\hf{\hat{f}}%

\global\long\def\hg{\hat{g}}%

\global\long\def\hh{\hat{h}}%

\global\long\def\hi{\hat{i}}%

\global\long\def\hj{\hat{j}}%

\global\long\def\hk{\hat{k}}%

\global\long\def\hl{\hat{l}}%

\global\long\def\hm{\hat{m}}%

\global\long\def\hn{\hat{n}}%

\global\long\def\ho{\hat{o}}%

\global\long\def\hp{\hat{p}}%

\global\long\def\hq{\hat{q}}%

\global\long\def\hr{\hat{r}}%

\global\long\def\hs{\hat{s}}%

\global\long\def\hu{\hat{u}}%

\global\long\def\hv{\hat{v}}%

\global\long\def\hw{\hat{w}}%

\global\long\def\hx{\hat{x}}%

\global\long\def\hy{\hat{y}}%

\global\long\def\hz{\hat{z}}%

\global\long\def\hA{\hat{A}}%

\global\long\def\hB{\hat{B}}%

\global\long\def\hC{\hat{C}}%

\global\long\def\hD{\hat{D}}%

\global\long\def\hE{\hat{E}}%

\global\long\def\hF{\hat{F}}%

\global\long\def\hG{\hat{G}}%

\global\long\def\hH{\hat{H}}%

\global\long\def\hI{\hat{I}}%

\global\long\def\hJ{\hat{J}}%

\global\long\def\hK{\hat{K}}%

\global\long\def\hL{\hat{L}}%

\global\long\def\hM{\hat{M}}%

\global\long\def\hN{\hat{N}}%

\global\long\def\hO{\hat{O}}%

\global\long\def\hP{\hat{P}}%

\global\long\def\hQ{\hat{Q}}%

\global\long\def\hR{\hat{R}}%

\global\long\def\hS{\hat{S}}%

\global\long\def\hT{\hat{T}}%

\global\long\def\hU{\hat{U}}%

\global\long\def\hV{\hat{V}}%

\global\long\def\hW{\hat{W}}%

\global\long\def\hX{\hat{X}}%

\global\long\def\hY{\hat{Y}}%

\global\long\def\hZ{\hat{Z}}%

\global\long\def\hap{\hat{\alpha}}%

\global\long\def\hbt{\hat{\beta}}%

\global\long\def\hgm{\hat{\gamma}}%

\global\long\def\hGm{\hat{\Gamma}}%

\global\long\def\hdt{\hat{\delta}}%

\global\long\def\hDt{\hat{\Delta}}%

\global\long\def\hep{\hat{\epsilon}}%

\global\long\def\hvep{\hat{\varepsilon}}%

\global\long\def\hzt{\hat{\zeta}}%

\global\long\def\het{\hat{\eta}}%

\global\long\def\hth{\hat{\theta}}%

\global\long\def\hvth{\hat{\vartheta}}%

\global\long\def\hTh{\hat{\Theta}}%

\global\long\def\hio{\hat{\iota}}%

\global\long\def\hkp{\hat{\kappa}}%

\global\long\def\hld{\hat{\lambda}}%

\global\long\def\hLd{\hat{\Lambda}}%

\global\long\def\hmu{\hat{\mu}}%

\global\long\def\hnu{\hat{\nu}}%

\global\long\def\hxi{\hat{\xi}}%

\global\long\def\hXi{\hat{\Xi}}%

\global\long\def\hpi{\hat{\pi}}%

\global\long\def\hPi{\hat{\Pi}}%

\global\long\def\hrh{\hat{\rho}}%

\global\long\def\hvrh{\hat{\varrho}}%

\global\long\def\hsg{\hat{\sigma}}%

\global\long\def\hSg{\hat{\Sigma}}%

\global\long\def\hta{\hat{\tau}}%

\global\long\def\hup{\hat{\upsilon}}%

\global\long\def\hUp{\hat{\Upsilon}}%

\global\long\def\hph{\hat{\phi}}%

\global\long\def\hvph{\hat{\varphi}}%

\global\long\def\hPh{\hat{\Phi}}%

\global\long\def\hch{\hat{\chi}}%

\global\long\def\hps{\hat{\psi}}%

\global\long\def\hPs{\hat{\Psi}}%

\global\long\def\hom{\hat{\omega}}%

\global\long\def\hOm{\hat{\Omega}}%

\global\long\def\hdgg#1{\hat{#1}^{\dagger}}%

\global\long\def\cjg#1{#1^{*}}%

\global\long\def\hsgx{\hat{\sigma}_{x}}%

\global\long\def\hsgy{\hat{\sigma}_{y}}%

\global\long\def\hsgz{\hat{\sigma}_{z}}%

\global\long\def\hsgp{\hat{\sigma}_{+}}%

\global\long\def\hsgm{\hat{\sigma}_{-}}%

\global\long\def\hsgpm{\hat{\sigma}_{\pm}}%

\global\long\def\hsgmp{\hat{\sigma}_{\mp}}%

\global\long\def\dert#1{\frac{d}{dt}#1}%

\global\long\def\dertt#1{\frac{d#1}{dt}}%

\global\long\def\Tr{\text{Tr}}%

\maketitle

\section{Introduction} \label{sec:intro}

Reservoir engineering is a powerful tool for quantum state synthesis in quantum technologies. In spite of the fact that the general idea has been around for about two decades~\cite{Cirac1993,MatosFilho1996,Poyatos1996,Luetkenhaus1998,Clark2003}, and experimental implementations have been reported since then~\cite{Myatt2000,Rabl2004,Kienzler2014,Lo2015}, there is still need for improvements, generalizations and new protocols~\cite{Woolley2014,Basilewitsch2019,Bai2021}. Recently, the stabilisation of a squeezed state using reservoir engineering was demonstrated with the squeezing factor greatly exceeding what is expected using other techniques~\cite{Dassonneville2021}. 

In~\cite{Roy2021}, reservoir engineering has been employed to steer entanglement and measurements in a composite system in order to guide an arbitrary initial state of the other part towards a chosen target state. The protocol requires multiple repetitions of an elementary step or ``collision'', which is itself an interesting tool for reservoir engineering. Previously, such repeated interactions have been used, for example, in the dissipative preparation of coherent states for the motional degree of freedom of  $^{40}$Ca$^+$ ions~\cite{Kienzler2014}, and the stabilization of non-classical states of radiation field in cavity quantum electrodynamics setups~\cite{Sarlette2011,Sarlette2012}.

Here, we will consider the framework of pulsed or collisional models~\cite{Ziman2002,Giovannetti2012} for engineering a multireservoir system for the motional state of a trapped ion. Multireservoir setups may enable the development of tasks such as the preparation of many-body states, non-equilibrium quantum phases and even  universal quantum computation without any coherent dynamics~\cite{Verstraete2009}. Our aim is to build on single reservoir engineering~\cite{Poyatos1996} and collisional models to expand them to multireservoir systems, which have a great theoretical~\cite{Cusumano2018} and experimental appeal~\cite{Kienzler2014}. 
In this paper, we also present its application in quantum engines, i.e., quantum systems undergoing either coherent (work) or incoherent (usually thermally) transformations induced in a series of steps bringing it back to its initial state~\cite{Scully2002,Quan2007}. After a cycle, the engine is expected to have converted part of the absorbed energy into net work. They usually operate far from equilibrium and are especially affected by quantum resources~\cite{Long2015,Camati2019,Das2020,Bresque2021,Medina2021} and particularities of their environments~\cite{Rossnagel2014,Cherubim2019,Camati2020}, making them particularly interesting to study quantum thermodynamics. 

The so called quantum Otto cycle (QOC) has become a paradigmatic example of a quantum engine investigated~\cite{Quan2007}. In particular, in the context of trapped ions, we find theoretical and experimental investigations of the QOC employing thermal~\cite{Abah2012,Rossnagel2016} and non-thermal baths~\cite{Rossnagel2014} with tapered trap geometries. These seminal works and the fact that trapped ions constitute a very mature experimental platform for quantum applications have motivated us to develop a multireservoir engineering protocol based on collisional interactions~\cite{Ciccarello2021}. 

Our protocol uses laser pulses on transitions in a single trapped ion to engineer its motional state. We illustrate our protocol on a trapped  $^{40}$Ca$^+$ calcium ion, which has suitable transitions for this scheme~\cite{Stute2012,Krutyanskiy2019,Takahashi2020}.

This paper is organized as follows. In section~\ref{sec:model}, we present the model describing the interaction of the multilevel trapped ion with pulsed lasers and the protocol to effectively create a multireservoir system in its vibrational degree of freedom. In sections~\ref{sec:appl} and~\ref{sec:otto}, we consider applications of our multireservoir protocol to quantum state synthesis and quantum engines. For the former, we consider the generation and protection of non-classical squeezed states, and the possibility of thermalization to arbitrary temperatures. For the latter, we will show how the open system dynamics with multiple reservoirs can be used to violate the Otto bound for the engine efficiency. In section~\ref{sec:conc}, we present our conclusions.

\section{Model} \label{sec:model}
\begin{figure*}[ht]
    \subfloat{\label{fig:modela}} 
	\subfloat{\label{fig:modelb}}
	\subfloat{\label{fig:modelc}}
	\subfloat{\label{fig:modeld}}
    \centering
    \includegraphics[width=\textwidth]{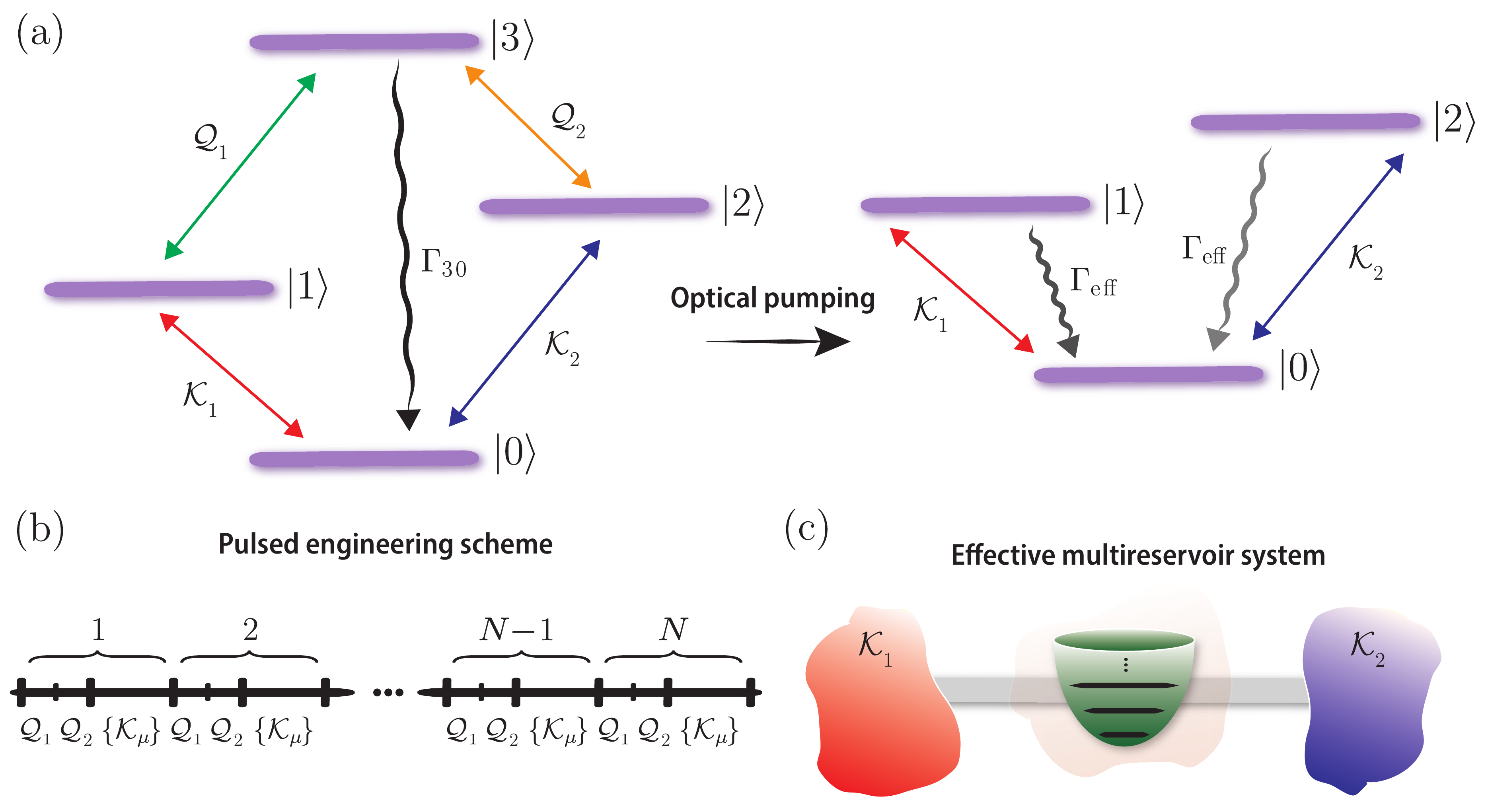}
    \caption{Scheme of the pulsed multireservoir engineering for a $d$--level trapped ion with $d=4$.}
    \label{fig:model}
\end{figure*}

We consider an ion of mass $M$ with $d$ internal levels trapped in a one-dimensional harmonic potential with angular frequency $\nu$. Its Hamiltonian is
\begin{align}
\hH_{\textrm{ion}}=\hbar\nu\left(\hdgg a\ha+\frac{1}{2}\right)+\hbar\sum_{\mu=0}^{d-1}\omega_{\mu}\kb{\mu}{\mu}, 
\label{eq:Hion}    
\end{align}
where $\hbar\omega_{\mu}$ is the energy of the internal level $\ket{\mu}$, and $\hdgg{a}$ ($\ha$) is the creation (annihilation) operator for the vibrational degree of freedom of the ion. The vibrational state of the ion is described by the Fock states $\{\ket{j_{\textrm{v}}}\}$. In our scheme, we choose the electronic levels $\{\ket{0},\ket{1},...,\ket{d-2}\}$ to have long lifetimes compared to the level $\ket{d-1}$, which spontaneously decays to the level $\ket{0}$ at a rate $\Gamma_{d-10}$ (figure~\ref{fig:modela}). This decay channel is considered to be much faster than any other incoherent process in the scheme, thus being the prominent source of decoherence in the model.

The reservoir engineering scheme we propose here then relies on the alternate action of  laser pulses addressing specific transitions of the ion as exemplified in the left panel of figure~\ref{fig:modela} and in figure~\ref{fig:modelb}. The lasers $\{\mathcal{K}_{\mu}\}$, hereafter also called \textit{engineering lasers}, are designed to be resonant with the transitions $\ket{0}\leftrightarrow\ket{\mu}$, $\mu=\{1,...,d-2\}$, and implement the open dynamics for the vibrational motion of the ion. The \textit{reset lasers} $\{\mathcal{Q}_{\mu}\}$ address the transitions $\ket{\mu}\leftrightarrow\ket{d-1}$, $\mu=\{1,...,d-2\}$, and they are responsible to create controlled decay channels for the levels $\{\ket{1},...,\ket{d-2}\}$, leading the internal degree of freedom of the ion to the ground state $\ket{0}$ by optical pumping (right panel of figure~\ref{fig:modela}). The lasers $\{\mathcal{Q}_{\mu}\}$ are applied in sequence to avoid undesirable couplings between the levels $\{\ket{1},...,\ket{d-2}\}$. 

While the choice $d=3$ leads to schemes of single reservoir engineering, implemented for instance in~\cite{Kienzler2014}, values of $d>3$ generate a multireservoir scenario so that one may simulate an effective interaction of the vibrational mode with $d-2$ independent Markovian baths (figure~\ref{fig:modelc}). In section~\ref{sec:appl}, we exemplify our protocol for $d=4$, show a speed-up in the generation of vibrational states of the ion, and demonstrate the synthesis of states that are unachievable with just a single reservoir. In particular, we demonstrate that the bireservoir case can be used to achieve thermalization to Gibbs states with arbitrary positive temperatures, which can correspond to either cooling or heating of the ion. 

The general description of the Hamiltonians for the ion-laser interactions in the rotating-wave approximation are given by  
\begin{align}
    \hH_{\mathcal{K}_{\mu}}(t)&=\hbar\Omega_{\text{r}}\sum_{j=1}^{n_\mu}\frac{\Omega_{\mu,j}}{\Omega_{\text{r}}}[e^{-i\omega_{\mu,j}t}e^{i\eta_{\mu,j}(\ha+\hdgg a)}\kb \mu0+\textrm{H.C.}],\label{eq:HK}\\
\hH_{\mathcal{Q}_{\mu}}(t)&=\hbar\tilde{\Omega}_{\text{r}}\sum_{j=1}^{\tilde{n}_\mu}\frac{\tilde{\Omega}_{\mu,j}}{\tilde{\Omega}_{\text{r}}}\left[e^{-i\tilde{\omega}_{\mu,j}t}e^{i\tilde{\eta}_{\mu,j}(\ha+\hdgg a)}\kb{d-1}{\mu}+\textrm{H.C.}\right],
\label{eq:HQ}
\end{align}
$\mu=\{1,...,d-2\}$. $\mathcal{K}_{\mu}$ ($\mathcal{Q}_{\mu}$) comprises of $n_{\mu}$ ($\tilde{n}_{\mu}$) lasers with angles $\phi_{\mu,j}$ ($\tilde{\phi}_{\mu,j}$) with respect to the trap axis, Rabi frequency $\Omega_{\mu,j}$ ($\tilde{\Omega}_{\mu,j}$) and angular frequency $\omega_{\mu,j}$ ($\tilde{\omega}_{\mu,j}$). The Lamb-Dicke parameters are given by $\eta_{\mu,j}=(\omega_{\mu,j}/c)\sqrt{\hbar/(2M\nu_{ })}\cos\phi_{\mu,j}$ ($\tilde{\eta}_{\mu,j}=(\tilde{\omega}_{\mu,j}/c)\sqrt{\hbar/(2M\nu_{ })}\cos\tilde{\phi}_{\mu,j}$) with the speed of light $c$. For convenience, we have written the Hamiltonians~\eqref{eq:HK} and~\eqref{eq:HQ} in terms of a reference Rabi frequencies $\Omega_{\textrm{r}}$ and $\tilde{\Omega}_{\textrm{r}}$, respectively. In the numerical simulations of section~\ref{sec:appl}, these are chosen as the Rabi frequencies of one of the lasers involved in the Hamiltonians.

In the interaction picture with respect to $\hH_{\text{ion}}$ and in the resolved sideband regime, the ion-laser Hamiltonians become time-independent provided that their angular frequencies fulfill the following conditions~\cite{Orszag2008} 
\begin{align}
\omega_{\mu,j}&=\omega_{\mu}-\omega_{0}-m_{\mu,j}\nu,\ \ \ \ \ \ \ \ \ \ \ \ \ \ \ \ \mu=\{1,...,d-2\}, \label{eq:sbrK}\\
\tilde{\omega}_{\mu,j}&=\omega_{d-1}-\omega_{\mu}-\tilde{m}_{\mu,j}\nu,\ \ \ \ \ \ \ \ \ \ \ \ \ \mu=\{1,...,d-2\}, \label{eq:sbrQ}
\end{align}
with $m_{\mu,j}$ and $\tilde{m}_{\mu,j}$ being integer numbers. In this case, the ion-laser Hamiltonians describe coherent phonon-excitation exchanges
\begin{align}
    \hH_{\mathcal{K}_{\mu}}&=\hbar\Omega_{\text{r}}\left(\hK_{\mu}\kb \mu0+\hdgg K_{\mu}\kb 0\mu\right), \label{eq:HK2}\\
\hH_{\mathcal{Q}_{\mu}}&=\hbar\tilde{\Omega}_{\text{r}}\left(\hQ_{\mu}\kb{d-1}{\mu}+\hdgg Q_{\mu}\kb{\mu}{d-1}\right),
\label{eq:HQ2}
\end{align}
where
\begin{align}
\hK_{\mu}=&\sum_{j=1}^{n_{\mu}}\frac{\Omega_{\mu,j}}{\Omega_{\text{r}}}\left\{ \frac{\left[1+\text{sgn}(m_{\mu,j})\right]}{2}\hd_{\n{m_{\mu,j}}} +\frac{\left[1+\text{sgn}(-m_{\mu,j})\right]}{2}\left(-1\right)^{\n{m_{\mu,j}}}\hdgg d_{\n{m_{\mu,j}}}\right\},\label{eq:K} \\
\hQ_{\mu}=&\sum_{j=1}^{\tilde{n}_{\mu}}\frac{\tilde{\Omega}_{\mu,j}}{\tilde{\Omega}_{\text{r}}}\left\{ \frac{\left[1+\text{sgn}(\tilde{m}_{\mu,j})\right]}{2}\hd_{\n{\tilde{m}_{\mu,j}}}+\frac{\left[1+\text{sgn}(-\tilde{m}_{\mu,j})\right]}{2}\left(-1\right)^{\n{\tilde{m}_{\mu,j}}}\hdgg d_{\n{\tilde{m}_{\mu,j}}}\right\}\label{eq:Q}
\end{align}
determine the different phonon exchange processes induced by the lasers according to $\{m_{\mu,j}\}$ and $\{\tilde{m}_{\mu,j}\}$, respectively with $\hd_{\n{m_{\mu,j}}}$ being
\begin{align}
\hd_{\n{m_{\mu,j}}}&=e^{-\eta_{\mu,j}^{2}/2}\sum_{k=0}^{\infty}\frac{\left(i\eta_{\mu,j}\right)^{2k+\n{m_{\mu,j}}}}{k!\left(k+\n{m_{\mu,j}}\right)!}\left(\hdgg a\right)^{k}\ha^{k}\ha^{\n{m_{\mu,j}}}, \label{eq:dK}\\
\hd_{\n{\tilde{m}_{\mu,j}}}&=e^{-\tilde{\eta}_{\mu,j}^{2}/2}\sum_{k=0}^{\infty}\frac{\left(i\tilde{\eta}_{\mu,j}\right)^{2k+\n{\tilde{m}_{\mu,j}}}}{k!\left(k+\n{\tilde{m}_{\mu,j}}\right)!}\left(\hdgg a\right)^{k}\ha^{k}\ha^{\n{\tilde{m}_{\mu,j}}}. \label{eq:dQ}
\end{align}

In the proposed scheme, we restrict the reset lasers to interact only with the internal states of the ion without disturbing its vibrational degree of freedom. This can be achieved, for instance, with lasers $\mathcal{Q}_{\mu}$ tuned to the carrier transition, which corresponds to $\tilde{m}_{\mu,1}=0$ in equation~\eqref{eq:sbrQ}. In the Lamb-Dicke regime ($\tilde{\eta}_{\mu,1}\ll1$), we can then rewrite equation~\eqref{eq:HQ2} 
\begin{align}
\hH_{\mathcal{Q}_{\mu}}\approx\hbar\tilde{\Omega}_{\text{r}}\left(\kb{d-1}{\mu}+\kb{\mu}{d-1}\right).
\label{eq:HQ3}
\end{align}
In the examples of section~\ref{sec:appl}, we also consider the Lamb-Dicke regime for the engineering lasers ($\eta_{\mu,j}\ll1$), which allows one to simplify the expression for $\hd_{\n{m_{\mu,j}}}$ in equation~\eqref{eq:dK}, by only considering two-phonon exchanges ($\n{m_{\mu,j}}=2$), one-phonon exchanges ($\n{m_{\mu,j}}=1$) and no-phonon exchanges ($\n{m_{\mu,j}}=0$). 

\subsection{Reset stages} \label{sec:reset}

The reset lasers $\{\mathcal{Q}_{\mu}\}$ create artificial decay channels for the long-lived levels $\{\ket{1},...,\ket{d-2}\}$ by inducing transitions to the short-lived level $\ket{d-1}$, as shown in figure~\ref{fig:modela} for $d=4$. This procedure is used to bring the electronic degree of freedom to the state $\ket{0}$ at the beginning of the protocol and between two consecutive engineering stages, where entanglement with the vibrational mode due to the interactions with lasers $\{\mathcal{K}_{\mu}\}$ is created.

A full reset cannot be achieved by the simultaneous action of all $\mathcal{Q}_{\mu}$ because of the creation of dark states between the $\ket{1},...\ket{d-2}$ states. The reset here is therefore carried out in $d-2$ steps, so that only a single $\mathcal{Q}_{\mu}$ is applied at a time, depopulating the level $\ket{\mu}$. In all steps, we assume that the Rabi frequency  $\tilde{\Omega}_{\textrm{r}}$ is much smaller than the spontaneous emission rate $\Gamma_{d-10}$, such that the level $\ket{d-1}$ can be adiabatically eliminated from the dynamics~\cite{Haroche2006}. 

Below, we exemplify this for $d=4$ and for the application of the laser $\mathcal{Q}_{1}$, whose Hamiltonian is given by equation~\eqref{eq:HQ3} with $\mu=1$. The spontaneous emission of the level $\ket{3}$ to the level $\ket{0}$ is described by the dissipator in Lindblad form
\begin{align}
\mathcal{D}_{30}(\hrh)=\Gamma_{30}\left[\kb 03\hrh\kb 30-\frac{1}{2}\left(\kb 33\hrh+\hrh\kb 33\right)\right],
\label{eq:diss}
\end{align}
where recoils effects on the vibrational motion of the ion are neglected, which is typically justified in the Lamb-Dicke regime and for a small number of phonons~\cite{Haroche2006}. The master equation for the total density operator $\hrh$ in the interaction picture is therefore given by
\begin{align}
    \dert{\hrh}=-\frac{i}{\hbar}[\hH_{\mathcal{Q}_1},\hrh]+\mathcal{D}_{30}(\hrh).
    \label{eq:ME2}
\end{align}
By defining the projections of $\hrh$ onto the electronic subspace as $\hrh_{jk}=\mel{j}{\hrh}{k}=\hrh_{kj}^{\dagger}$, from equation~\eqref{eq:ME2} we readily obtain $d\hrh_{02}/dt= d\hrh_{20}/dt=d\hrh_{22}/dt=0$, and 
\begin{align}
\dert{\hrh_{00}} &=\Gamma_{30}\hrh_{33},\ \ \ 
\dert{\hrh_{11}} =-i\tilde{\Omega}_{\text{r}}\left(\hrh_{31}-\hrh_{13}\right),\ \ \ \dert{\hrh_{01}}=i\tilde{\Omega}_{\text{r}}\hrh_{03},\ \ \ \dert{\hrh_{12}} =-i\tilde{\Omega}_{\text{r}}\hrh_{32}, \nonumber \\
\dert{\hrh{}_{33}}&=-i\tilde{\Omega}_{\text{r}}\left(\hrh_{13}-\hrh_{31}\right)-\Gamma_{30}\hrh_{33},\ \ \ 
\dert{\hrh_{03}} =i\tilde{\Omega}_{\text{r}}\hrh_{01}-\frac{\Gamma_{30}}{2}\hrh_{03},\nonumber \\
\dert{\hrh_{13}} &=-i\tilde{\Omega}_{\text{r}}\left(\hrh_{33}-\hrh_{11}\right)-\frac{\Gamma_{30}}{2}\hrh_{13},\ \ \ 
\dert{\hrh_{23}} =i\tilde{\Omega}_{\text{r}}\hrh_{21}-\frac{\Gamma_{30}}{2}\hrh_{23}.
\label{eq:opump}
\end{align}
The condition $\tilde{\Omega}_{\textrm{r}}/\Gamma_{30}\ll1$ allows one to write
$d\hrh_{33}/dt\approx d\hrh_{23}/dt\approx d\hrh_{13}/dt\approx d\hrh_{03}/dt\approx0$ in the set of equations~\eqref{eq:opump}. With $\hrh_{33}\ll\hrh_{11}$, one finds that $\hrh_{11}$ goes to zero at an effective rate $\Gamma_{\textrm{eff}}\equiv4(\tilde{\Omega}_{\textrm{r}})^2/\Gamma_{30}$ while $\hrh_{22}$ stays basically constant. Similarly, the subsequent application of $\mathcal{Q}_2$ makes $\hrh_{22}$ go to zero at the same rate. The overall effect of the two-step reset stage is to asymptotically drive $\hrh_{00}$ to unity, as desired, while having an adjustable rate $\Gamma_{\textrm{eff}}$. Consequently, one can estimate the total time of a $(d-2)$--step reset stage as being $t_{\text{reset}}\propto (d-2)\Gamma_{\textrm{eff}}^{-1}$.

\subsection{Engineering stages}

The engineering stages of the pulsed scheme rely only on the unitary evolution induced by the lasers $\{\mathcal{K}_{\mu}\}$, while $\hH_{\mathcal{Q}_\mu}=0$. This allows one to treat the electronic degree of freedom of the ion as an effective ($d-1$)-level system with the corresponding basis set $\{\ket{0},...,\ket{d-2}\}$.
For a single engineering stage of duration $\tau_{\textrm{r}}$, the time-evolved density operator for the vibrational part of the ion $\hrh_{\textrm{v}}(\tau_{\textrm{r}})$, in an initially separable state, can be expressed as
\begin{align}
\hrh_{\textrm{v}}(\tau_{\textrm{r}})=\Tr_{\textrm{e}}\left[\hU_{\mathcal{K}}(\tau_{\textrm{r}})\hrh_{\textrm{v}}(0)\hrh_{\textrm{e}}(0)\hdgg{U}_{\mathcal{K}}(\tau_{\textrm{r}})\right]
,\label{eq:rhov}
\end{align}
where the trace is taken over the electronic levels, $\hU_{\mathcal{K}}\left(\tau_{\textrm{r}}\right)=e^{-i\hH_{\mathcal{K}}\tau_{\textrm{r}}/\hbar}$ is the unitary time-evolution operator associated with the engineering lasers, and
\begin{align}
\hH_{\mathcal{K}}&=\hbar\Omega_{\text{r}}\sum_{\mu=1}^{d-2}\left(\hK_{\mu}\kb{\mu}{0}+\hdgg K_{\mu}\kb{0}{\mu}\right), \label{eq:HK3}
\end{align}
with $\hK_{\mu}$ given by equation~\eqref{eq:K}. 

The second order expansion of the time-evolution operator is
\begin{align}
\hU_{\mathcal{K}}\left(\tau_{\textrm{r}}\right)\approx \hI-\frac{i\tau_{\textrm{r}}}{\hbar}\hH_{\mathcal{K}}-\frac{1}{2}\left(\frac{\tau_{\textrm{r}}}{\hbar}\right)^2\hH_{\mathcal{K}}^2. 
\label{eq:U2}    
\end{align}
Retaining all terms up to second order in equation~\eqref{eq:rhov} results in
\begin{align}
\hrh_{\textrm{v}}(\tau_{\textrm{r}})\approx\  &\Tr_{\textrm{e}}\left[\hrh_{\textrm{v}}(0)\hrh_{\textrm{e}}(0)\right]+\frac{i\tau_{\textrm{r}}}{\hbar}\left\{ \Tr_{\textrm{e}}\left[\hrh_{\textrm{v}}(0)\hrh_{\textrm{e}}(0)\hH_{\mathcal{K}}\right]-\Tr_{\textrm{e}}\left[\hH_{\mathcal{K}}\hrh_{\textrm{v}}(0)\hrh_{\textrm{e}}(0)\right]\right\}\nonumber \\
{}&+\left(\frac{\tau_{\textrm{r}}}{\hbar}\right)^{2}\left\{ \Tr_{\textrm{e}}\left[\hH_{\mathcal{K}}\hrh_{\textrm{v}}(0)\hrh_{\textrm{e}}(0)\hH_{\mathcal{K}}\right]-\frac{1}{2}\Tr_{\textrm{e}}\left[\hH_{\mathcal{K}}^{2}\hrh_{\textrm{v}}(0)\hrh_{\textrm{e}}(0)\right]-\frac{1}{2}\Tr_{\textrm{e}}\left[\hrh_{\textrm{v}}(0)\hrh_{\textrm{e}}(0)\hH_{\mathcal{K}}^{2}\right]\right\}. \label{eq:rhov2}
\end{align}
With the particular choice of $\hrh_{\textrm{e}}(0)=\kb 00$, as provided by the reset stages, the linear term in $\tau_{\textrm{r}}$ in equation~\eqref{eq:rhov2} vanishes, such that we are left with
\begin{align}
    \hrh_{\textrm{v}}(\tau_{\textrm{r}})\approx\sum_{j=0}^{d-2}\hM_{j}(\tau_{\textrm{r}})\hrh_{\textrm{v}}(0)\hdgg M_{j}(\tau_{\textrm{r}}),\label{eq:rhov3}
\end{align}
where we have defined the Kraus-like operators
\begin{align}
\hM_{0}(\tau_{\textrm{r}})&=\hI-\frac{\left(\Omega_{\textrm{r}}\tau_{\textrm{r}}\right)^{2}}{2}\sum_{\mu=1}^{d-2}\hdgg K_{\mu}\hK_{\mu},\nonumber \\ \hM_{\mu}(\tau_{\textrm{r}})&=-i\Omega_{\textrm{r}}\tau_{\textrm{r}}\hK_{\mu},\ \ \ \ \ \mu=\{1,...,d-2\},\label{eq:Mop}
\end{align}
which satisfy $\sum_{j=0}^{d-2}\hM_{j}^{\dagger}(\tau_{\textrm{r}})\hM_{j}(\tau_{\textrm{r}})=\hI$ up to $(\Omega_{\textrm{r}}\tau_{\textrm{r}})^2\hdgg K_{\mu}\hK_{\mu}$. In other words, the vibrational state of the ion after an engineering stage of duration $\tau_{\textrm{r}}$ can be written in terms of the quantum map $\Phi_{\tau_{\textrm{r}}}(\cdot)=\sum_{j=0}^{d-2}\hM_{j}(\tau_{\textrm{r}})(\cdot)\hdgg M_{j}(\tau_{\textrm{r}})$, which in turn has the semigroup property $\Phi_{\tau_{\textrm{r}}}[\Phi_{\tau_{\textrm{r}}}(\cdot)]=\Phi_{2\tau_{\textrm{r}}}(\cdot)$ when a second order expansion is performed in equation~\eqref{eq:rhov}. This suggests that the fast consecutive applications of the engineering lasers interposed by the reset stages tend to reproduce a Markovian dynamics for the vibrational motion of the ion. Indeed, after the $N$--th interaction with the lasers $\{\mathcal{K}_\mu\}$, we can to write $\hrh_{\textrm{v}}\left(N\tau_{\textrm{r}}\right)\equiv \hrh_{N}$ as
\begin{equation}
\hrh_{N}=\hrh_{N-1}+\left(\Omega_{\textrm{r}}\tau_{\textrm{r}}\right)^{2}\sum_{\mu=1}^{d-2}\left[\hK_{\mu}\hrh_{N-1}\hdgg K_{\mu}-\frac{1}{2}\left\{ \hdgg K_{\mu}\hK_{\mu},\hrh_{N-1}\right\} \right],
\label{eq:rhov4}
\end{equation}
so that the final vibrational state depends only on its state at the end of the previous engineering stage. 

The main result of this Section is thus expressed by equation~\eqref{eq:rhov4}, since it provides a recursive and numerical recipe for the implementation of the Markovian open dynamics with operators $\hK_{\mu}$ chosen according to equation~\eqref{eq:K}.
Dividing $\Delta \hrh_{\textrm{v}}\equiv \hrh_{N}-\hrh_{N-1}$ by a time interval $\Delta t$, assuming $\Delta t \approx \tau_{\textrm{r}}$, taking the limit $\Delta t\rightarrow 0$ (or equivalently, $N\rightarrow\infty$) and introducing $\gamma = \Omega_{\textrm{r}}^2\/\tau_{\textrm{r}}$ the multireservoir master equation can be written in the Lindblad form
\begin{align}
\dert{\hrh_\textrm{v}}=\gamma\sum_{\mu=1}^{d-2} \left( \hK_{\mu} \hrh_{\textrm{v}} \hdgg K_{\mu}-\frac{1}{2}\left\{ \hdgg K_{\mu} \hK_{\mu},\hrh_{\textrm{v}}\right\} \right).
\label{eq:ME}
\end{align}

For the subsequent numerical calculations involving $\hrh_N$, we truncate the vibrational Hilbert space at a sufficiently excited number state $\ket{j_{\textrm{v}}^{\textrm{max}}}$, and use the semigroup property of the effective dynamical map to write
\begin{align}
    \mathbf{\rho}_N=e^{N\hL}\mathbf{\rho}_0, \label{eq:rhoNvec}
\end{align}
with $\mathbf{\rho}_N$ and $\mathbf{\rho}_0$ being the respective vectorized forms of $\hrh_N$ and $\hrh_0$~\cite{gilchrist2011}, and 
\begin{equation}
\hat{L}=\sum_{\mu=1}^{d-2} \epsilon_{\mu}\left\{ \hK_{\mu}^{\prime *} \otimes\hK^{\prime}_{\mu}-\frac{1}{2}\left[\hI\otimes\hat{K}^{\dagger \prime}_{\mu} \hK^{\prime}_{\mu}+\left(\hat{K}^{\dagger \prime}_{\mu} \hK^{\prime}_{\mu}\right)^{\top} \otimes\hI\right]\right\},
\label{eq:Liouv} 
\end{equation}
where $\epsilon_{\mu}$ and $\hK'_{\mu}$ emerge from a rescaling of $\hK_{\mu}$. This rescaling is convenient as it eliminates constant factors from one of the terms arising in the $\hd_{\n{m_{\mu,j}}}$ operators defined in equation~\eqref{eq:dK}, rendering $\epsilon_{\mu}$ the role of damping coefficients controlling the weight of each produced reservoir.

\section{Quantum state synthesis and protection}
\begin{figure}[ht]
\begin{center}
\subfloat{\label{fig:levelsca40a}} 
\subfloat{\label{fig:levelsca40b}}
\includegraphics[width=1\linewidth]{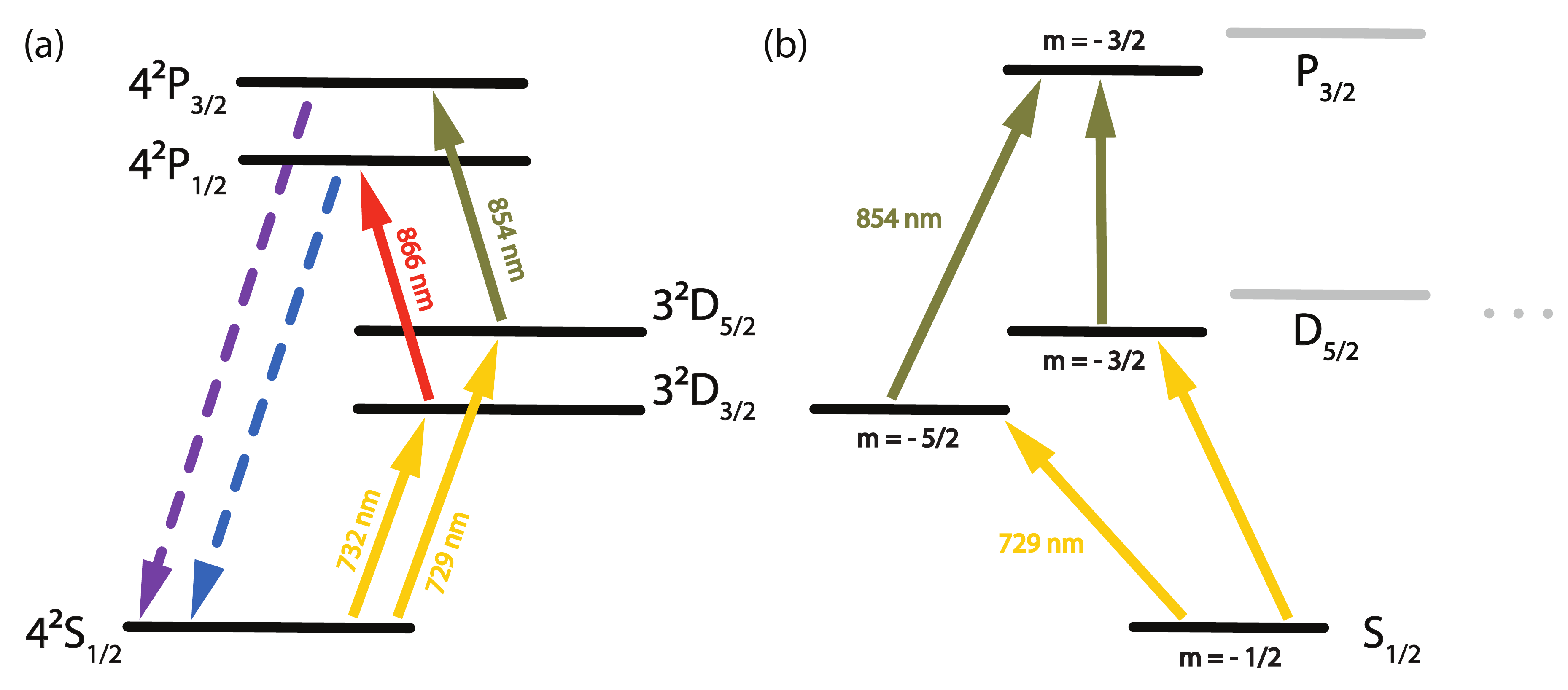} 
\caption{Relevant transitions for the implementation of two-reservoir engineering with a $^{40}$Ca$^+$ trapped ion~\cite{Takahashi2020,Ross1999} using (a) the ion's fine structure or (b) the Zeeman structure of the $D_{5/2}$--level.}
\label{fig:levelsca40} 
\end{center}
\end{figure}

{\label{sec:appl}}
In order to illustrate our formalism, we present example applications in the simplest scenario with $d=4$. This will create two independent reservoirs  (figure~\ref{fig:modelc}). 
For example, such a system can be implemented with trapped $^{40}$Ca$^+$ ions in two ways. The relevant energy levels and driving lasers are depicted in figure~\ref{fig:levelsca40}. Quadruple transitions $S_{1/2}\leftrightarrow D_{3/2}$ and $S_{1/2}\leftrightarrow D_{5/2}$ are used during the engineering stage, while $D_{3/2}\leftrightarrow P_{1/2}$ and $D_{5/2}\leftrightarrow P_{3/2}$ allows for the reset to $S_{1/2}$ as discussed in section~\ref{sec:reset} (see figure~\ref{fig:levelsca40a}). 
Alternatively, the system can be created by exploiting the Zeeman structure of the $D_{5/2}$ state in conjunction with a magnetic field to lift the level degeneracy (see figure~\ref{fig:levelsca40b}). As described in section 2, by choosing the motional sideband transitions ${m_{\mu,j}}$ and the engineering laser interaction, a wide variety of motional states can be created.

In Table~\ref{tb:Kprime}, we show examples of possible choices of integers $\{m_{\mu,j}\}$ and re-scaled operators $\hK'_{\mu}$, which lead to the vibrational target state $\hrh_{\textrm{v}}^{\textrm{ref}}=\kb{\psi_{\textrm{v}}^{\textrm{ref}}}{\psi_{\textrm{v}}^{\textrm{ref}}}$ when all $\hK'_{\mu}$ operators are the same. In the table, we assume that $\{\eta_{\mu,j}\}\approx \eta \ll 1$ and set up $\Omega_{\textrm{r}}=\Omega_{\mu,{1}}$, which yields the increments 
\begin{align}
    \epsilon_{\mu}=\Omega_{\mu,1}^2\tau_{\textrm{r}}^2\eta^2 e^{-\eta^2}. \label{eq:eps}
\end{align}
The protocol can generate, for instance, (i) \textbf{coherent states} and (ii) \textbf{squeezed states}, with mode displacement and squeezing operators defined respectively as
\begin{align} 
\hD(\alpha) & =\exp\left(\alpha\ha^{\dagger}-\alpha^{*}\ha\right),\label{eq:Dopgen}\\ 
\hS(r) & =\exp\left[\frac{r}{2}\left(\ha^{\dagger2}-\ha^{2}\right)\right],\label{eq:Sopgen}
\end{align}
where $\alpha$ is the complex displacement in phase space and $r$ is the real-valued squeezing parameter. The combination of displacement and squeezing yields (iii) \textbf{squeezed coherent states}, whose application in quantum Otto cycles will be discussed in section~\ref{sec:otto}. According with the choices of integers $\{m_{\mu,j}\}$ in Table~\ref{tb:Kprime}, the displacements $\alpha_{\mu,\text{coh}}$, $\alpha_{\mu,\text{sqcoh}}$, and the squeezing parameter $r_{\mu}$ are given by
\begin{align}
    \alpha_{\mu,\text{coh}}&=i \frac{\Omega_{\mu,2}}{\Omega_{\mu,1}\eta},\ \ \ 
    \alpha_{\mu,\text{sqcoh}}=i \frac{\Omega_{\mu,3}}{\Omega_{\mu,1}\eta},\ \ \ 
    r_{\mu}=\tanh^{-1}\left(\frac{\Omega_{\mu,2}}{\Omega_{\mu,1}}\right),\label{eq:param}
\end{align}
with $\mu\in\{1,2\}$. 
While (i)-(iii) can also be generated in a single-bath configuration~\cite{Cirac1993,Kienzler2014}, our approach for (iii) is more efficient (see below). However, the incoherent combination of cases (iv) and (v) which can be used to synthesize \textbf{thermal states}, an important ingredient in quantum thermodynamics, may only be achieved using a multireservoir setup. A single bath configuration with $\hK'_{\mu}=\ha+\hdgg{a}$ does not produce a thermal state asymptotically.

In our protocol, the choices $m_{1,1}=1$ and $m_{2,1}=-1$ mimic the thermalization master equation for the vibrational mode, and a positive temperature can be ascribed to the state if $\epsilon_{1}>\epsilon_{2}$. By choosing the Lamb-Dicke parameters as $\eta_{1,1}=\eta_{2,1}=\eta$, the mean occupation number of the produced state becomes
\begin{equation} \bar{n}=\left[\left(\frac{\Omega_{1,1}}{\Omega_{2,1}}\right)^{2}-1\right]^{-1},\label{eq:renbar} \end{equation} 
with $\Omega_{1,1}>\Omega_{2,1}$. Therefore, such choices can promote either cooling or heating of the ion in the sideband regime without resorting to knowledge of the internal electronic dynamics through its correlation functions, thus being distinct from the scheme of Ref.~\cite{Cirac1992}.

\begin{table*}
\caption{Selected choices of integer numbers $\{m_{\mu,j}\}$ producing the effective laser operators $\hK'_{\mu}$ and the final state $\ket{\psi_{\textrm{v}}^{\textrm{ref}}}$.}
\begin{centering}
\begin{tabular}{p{0.3cm} p{5.6cm}|p{4.4cm}|p{4cm}}
\hline \hline 
$\{m_{\mu,j}\}$ & {} & $\hK'_{\mu}$& $\ket{\psi_{\textrm{v}}^{\textrm{ref}}}$\tabularnewline \hline \hline
(i) & $m_{\mu,1}=1$; $m_{\mu,2}=0$ & $\ha-\alpha_{\mu,\textrm{coh}}\hI$ & $\hD(\alpha_{\mu,\textrm{coh}})\ket{0_{\textrm{v}}}$ \tabularnewline
(ii) & $m_{\mu,1}=1$; $m_{\mu,2}=-1$ & $\ha+\tanh{(r_{\mu})}\hdgg{a}$ & $\hS(-r_{\mu})\ket{0_{\textrm{v}}}$\tabularnewline
(iii) & $m_{\mu,1}=1$; $m_{\mu,2}=-1$; $m_{\mu,3}=0$ & $\ha+\tanh{(r_{\mu})}\hdgg{a}-\alpha_{\mu,\textrm{sqcoh}}\hI$ & $\hS(-r_{\mu})\hD(\alpha_{\mu,\textrm{sqcoh}})\ket{0_{\textrm{v}}}$\tabularnewline
(iv) & $m_{\mu,1}=1$ & $\ha$ & $\ket{0_{\textrm{v}}}$ \tabularnewline
(v) & $m_{\mu,1}=-1$ & $\hdgg{a}$ & $\ket{j_{\textrm{v}}^{\textrm{max}}}$
\tabularnewline
\hline \hline
\end{tabular}
\par\end{centering}
\label{tb:Kprime}
\end{table*}

For the simulations, the performance of the method is numerically investigated with the help of the fidelities between quantum states~\cite{Jozsa1994}
\begin{align}
\mathcal{F}_{\infty}&=\left(\Tr\sqrt{\sqrt{\hrh_{\textrm{v}}^{\textrm{ref}}}\hrh_{N}\sqrt{\hrh_{\textrm{v}}^{\textrm{ref}}}}\right)^{2},\label{eq:fidsteady}\\
\mathcal{F}_{0}&=\left(\Tr\sqrt{\sqrt{\hrh_{0}}\hrh_{N}\sqrt{\hrh_{0}}}\right)^{2}.\label{eq:fid0}
\end{align}
The fidelity $\mathcal{F}_{\infty}$ indicates the distance between the vibrational state produced after the $N$--th engineering stage, $\hrh_{N}$ given by equation~\eqref{eq:rhov4}, and the steady state associated with the target dynamics, $\hrh_{\textrm{v}}^{\textrm{ref}}$, which is obtained from equation~\eqref{eq:ME} by imposing $d\hrh_{\textrm{v}}^{\textrm{ref}}/dt=0$. On the other hand, the fidelity $\mathcal{F}_{0}$ indicates the distance between the state generated by the pulsed protocol and a particular initial state $\hrh_0$. Therefore, while $\mathcal{F}_{\infty}$ measures the protocol's success to synthesis the target state, $\mathcal{F}_{0}$ shows how different incoherent processes affect the vibrational mode, which allows for the study the state protection through reservoir engineering. We use the vectorization procedure described in section~\ref{sec:model} to obtain the density operators in equations~\eqref{eq:fidsteady} and~\eqref{eq:fid0}.

\begin{figure}[ht]
\begin{center}
\subfloat{\label{fig:fidNottoa}} 
\subfloat{\label{fig:fidNottob}}
\subfloat{\label{fig:fidNottoc}} 
\subfloat{\label{fig:fidNottod}}
\includegraphics[width=0.9\linewidth]{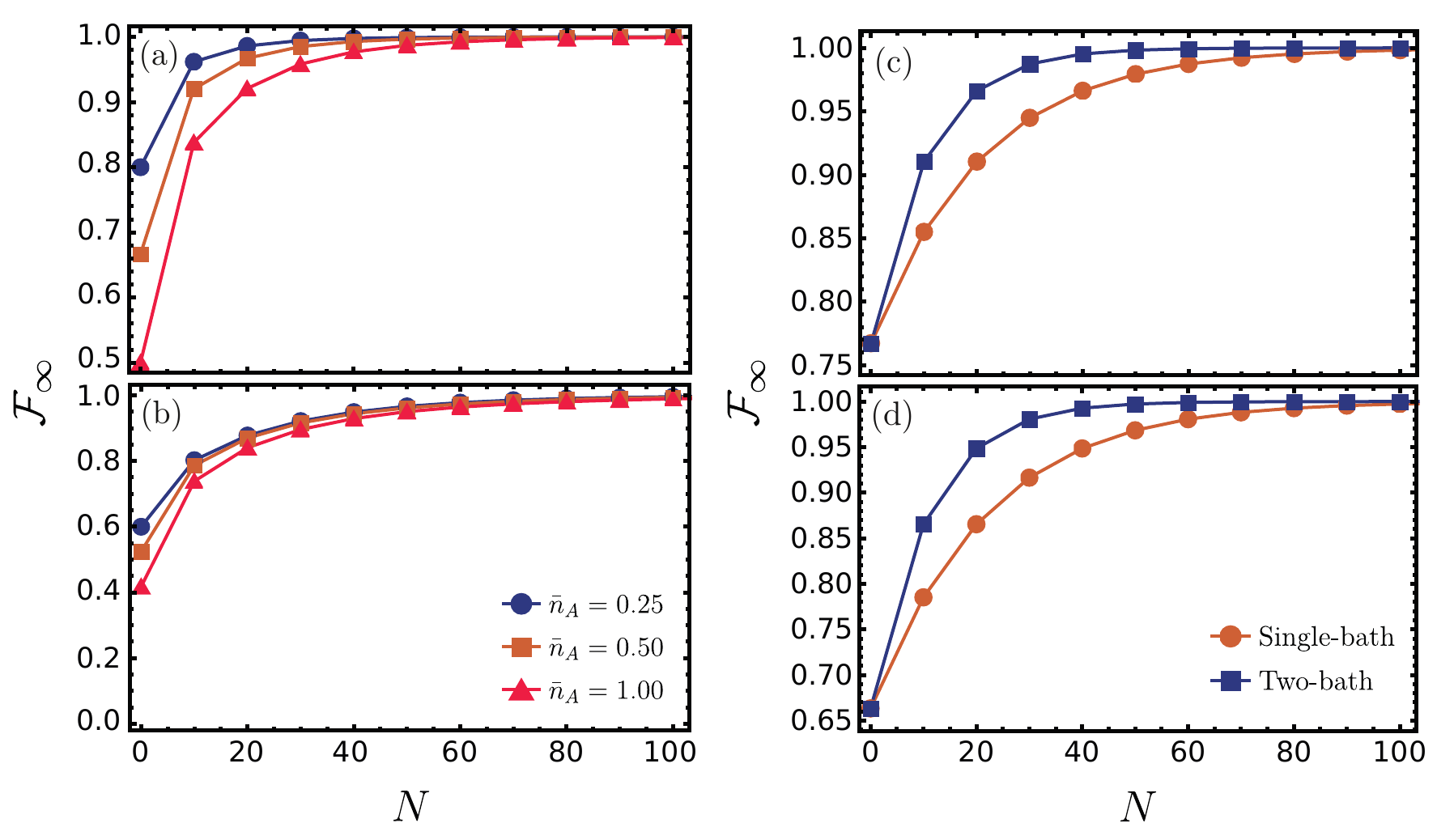} 
\caption{Vibrational state synthesis through the pulsed multireservoir engineering for a $4$--level trapped ion. The plots show the fidelities $\mathcal{F}_{\infty}$, defined in equation~\eqref{eq:fidsteady}, as functions of the number of engineering stages $N$. The \textbf{left panels} show the generation of thermal states for selected mean occupation numbers $\bar{n}_A$ such that  $\Omega_{1,1}/\Omega_{2,1}=5.0$ (circle markers); $\Omega_{1,1}/\Omega_{2,1}=3.0$ (square markers); $\Omega_{1,1}/\Omega_{2,1}=2.0$ (triangle markers). The chosen initial states are: (a) $\ket{0_{\text{v}}}$ and (b) $\hD(0.60i)\ket{0_{\textrm{v}}}$. The \textbf{right panels} show the generation of squeezed coherent states from the ground state $\ket{0_{\text{v}}}$ in the single-bath (circle markers) and the two-bath (square markers) configurations. In these cases, the generated states have squeezing parameter $r_{1}=0.11$ and displacements (c) $\alpha_{1,\textrm{sqcoh}}=0.48i$ and (d) $\alpha_{1,\textrm{sqcoh}}=0.60i$. The remaining parameters are chosen as $\eta=5.0\times 10^{-2}$, $\Omega_{\textrm{r}}\tau_{\textrm{r}}=4.5$, $\Omega_{1,1}/\Omega_{\textrm{r}}=1.0$, $\Omega_{1,2}/\Omega_{\textrm{r}}=1.1\times 10^{-1}$, $\Omega_{1,3}/\Omega_{\textrm{r}}=2.4\times 10^{-2}$ [case (c)] and $\Omega_{1,3}/\Omega_{\textrm{r}}=3.0\times 10^{-2}$ [case (d)].}
\label{fig:fidNotto} 
\end{center}
\end{figure}

Figure~\ref{fig:fidNotto} shows the fidelities $\mathcal{F}_{\infty}$ as functions of the number $N$ of engineering stages for a fixed value of the Lamb-Dicke parameter $\eta=5.0\times 10^{-2}$. The left panels in figure~\ref{fig:fidNotto} refer to the generation of thermal states of the vibrational mode starting from two paradigmatic examples of initial conditions: a ground state (figure~\ref{fig:fidNottoa}) and a coherent state (figure~\ref{fig:fidNottob}). As mentioned before, this is beyond what is possible with canonical single reservoir methods as it requires the incoherent superposition of multiple baths using, for instance, the method developed here. Our approach allows also for the control of the thermal mean occupation number by adjusting the relative intensities of the lasers as given by equation~\eqref{eq:renbar}. Starting from the ground state $\ket{0_{\textrm{v}}}$ (figure~\ref{fig:fidNottoa}), fidelities over $95\%$ are produced with $N\sim 10$ ($\bar{n}_A=0.25$), $N\sim 20$ ($\bar{n}_A=0.50$) and $N\sim 40$ ($\bar{n}_A=1.0$). Starting from the coherent state $\hD(0.60i)\ket{0_{\textrm{v}}}$ (figure~\ref{fig:fidNottob}), fidelities of $>95\%$ are reached with $N\sim 60$. The increase in the required number of engineering stages compared to the ion initially in the ground state reflects the presence of coherences in the initial state.

The right panels of figure~\ref{fig:fidNotto} shows the synthesis of squeezed coherent states for   single and two-bath protocols. The vibrational mode is initially in the ground state and the Rabi frequencies are chosen to yield the squeezing parameter $r_{\mu}=0.11\approx (\Omega_{\mu,2}/\Omega_{\mu,1})$. Different displacements $\alpha_{\mu,\text{sqcoh}}$ are produced in figures.~\ref{fig:fidNottoc} and \ref{fig:fidNottod} through the control of the third Rabi frequency $\Omega_{\mu,3}$ as given in equations~\eqref{eq:param}. The fidelity $\mathcal{F}_{\infty}$ for the two-bath protocol surpasses $95\%$ already for $N\sim 20$ (squared markers) while the single-bath protocol requires $N\sim 40$ (circle markers), indicating a significantly faster state synthesis. This is a natural consequence of the enhanced damping rate by including the second effective reservoir. Fast state synthesis is advantageous in quantum information protocols due to the inevitable sources of decoherence in the experimental setup. For quantum engines, fast dissipative processes can also help to reduce the total cycle time, thus increasing their total power. In addition, it is important to highlight that such a faster state synthesis occurs without resorting to stronger laser intensities, which might be challenging to obtain experimentally or also breakdown the rotating-wave approximation in equation~\eqref{eq:HK}. However, due to the sequential aspect of each of the reset stages, the speed of the state generation described in our protocol is physically limited by the reset mechanism, which in turn becomes more involved for very large $d$.

\begin{figure}[ht]
\begin{center}
\includegraphics[width=0.55\linewidth]{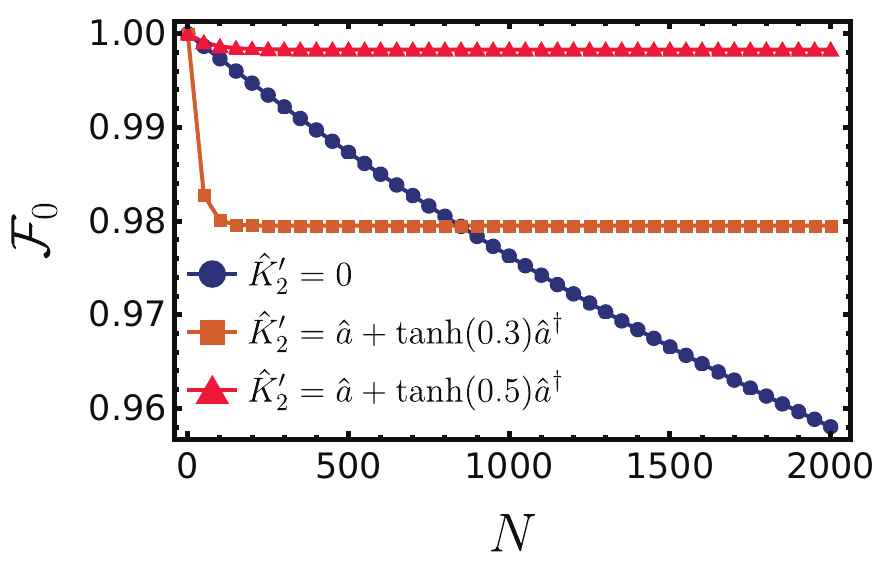} 
\caption{Fidelity between $\hrh_0=\hS(-0.50)\kb{0_{\textrm{v}}}{0_{\textrm{v}}}\hS(0.50)$ and $\hrh_N$ as a function of $N$ with the choices $\hK'_1=\ha$ and $\epsilon_1=1.0\times 10^{-4}$. While the circle markers show the degradation of $\hrh_0$ solely attributed to $\hK'_1$, the square and triangle markers show the inclusion of a second bath with the choices $\hK'_2=\ha+\tanh(r_2)\hdgg{a}$ and $\epsilon_2=2.0\times 10^{-2}$. The remaining parameters are $\eta=5.0\times 10^{-2}$, $\Omega_{\textrm{r}}\tau_{\textrm{r}}=2.0\times 10^{-1}$, $\Omega_{1,1}/\Omega_{\textrm{r}}=1.0$, $\Omega_{2,1}/\Omega_{\textrm{r}}=14$, $\Omega_{2,2}/\Omega_{\textrm{r}}=4.1$ (square markers) and $\Omega_{2,2}/\Omega_{\textrm{r}}=6.5$ (triangle markers).}
\label{fig:fid0} 
\end{center}
\end{figure}

 The simulation of competing dynamics provides another instance where the use of a engineered multireservoir setup might be useful. In this respect, we will now consider the generation of squeezed states $\ket{\psi_0}=\hS(-r)\ket{0_{\textrm{v}}}$ as the result of the interaction of a quantum harmonic oscillator and a reservoir which induces $\hK'_2=\ha+\tanh(r_2)\hdgg{a}$. This mechanism will compete with dissipation $\hK'_1=\ha$ which is ubiquitous all over practical implementations of resonators. In figure~\ref{fig:fid0}, where we plot $\mathcal{F}_0$ as a function of $N$, we clearly see the effect of the competition of these two mechanisms whose features are externally controlled by means of the applied lasers. The fidelity $\mathcal{F}_0$ is evaluated with respect to $\hrh_0=\ket{\psi_0}\bra{\psi_0}$ with a fixed value of the compression parameter $r$. One can also look at this plot having another physical application in mind. Let us suppose that the state $\ket{\psi_0}=\hS(-r)\ket{0_{\textrm{v}}}$ has been prepared at $t=0$. Figure~\ref{fig:fid0} reveals that a carefully engineered bath can inhibit the detrimental action of dissipation. 

\section{Generalized quantum Otto cycles with a trapped ion}{\label{sec:otto}}

In this section we discuss the use of the pulsed multireservoir method to implement a quantum Otto cycle (QOC) operating as a heat engine. Quantum heat engines are a powerful tool to investigate quantum effects on the energy exchange processes with trapped ions being a promising technology for experimental tests. In particular, QOCs also provide the possibility to easily distinguish unitary (coherent) from non-unitary (incoherent) changes of mean energy of the system, usually referred to as work and heat, respectively. The pulsed multireservoir engineering used during these non-unitary steps allows the practical investigations beyond purely thermal reservoirs.

\subsection{General definitions}\label{sec:def}

\begin{figure}[ht]
\begin{center}
\includegraphics[width=0.45\linewidth]{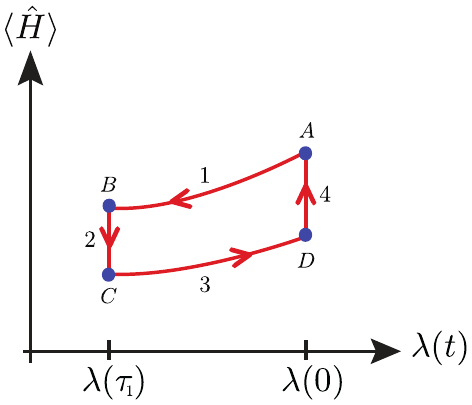}
\caption{Schematic representation of a quantum Otto cycle in the parameter--mean energy space for a single control parameter $\lambda(t)$.}
\label{fig:ottocycle} 
\end{center}
\end{figure}

The work substance is described by a time-dependent Hamiltonian $\hH(t)$ and density operator $\hrh(t)$ at an instant $t$. The time-dependence of $\hH(t)$ is promoted by the control of a set of parameters $\{\lambda_i(t)\}$. The QOC then connects the points $\mathbf{A}$, $\mathbf{B}$, $\mathbf{C}$, and $\mathbf{D}$ in the parameter--mean energy space through two unitary processes (strokes $1$ and $3$), and two non-unitary processes (strokes $2$ and $4$), each with duration $\tau_j$, $j=\{1,2,3,4\}$ (figure~\ref{fig:ottocycle}). 
The system is initialized in the state $\hrh(0)=\hrh_A$ and during the first stroke its Hamiltonian is changed from $\hH(0)$ to $\hH(\tau_1)$ through the unitary time-evolution operator $\hU(\tau_1)=\mathcal{T}\exp\left[-\frac{i}{\hbar}\int_{0}^{\tau_1}dt\hH(t)\right]$. In the second stroke, the system's Hamiltonian is kept fixed at $\hH(\tau_1)$ and an interaction with a reservoir is turned on leading the system to state $\hrh(\tau_1+\tau_2)=\hrh_C$. The third stroke comprises the change of Hamiltonian from $\hH(\tau_1)$ back to $\hH(0)$ through $\hU^{\dagger}(\tau_1)$, so that $\tau_3=\tau_1$. Finally, during the fourth stroke, $\hH(0)$ is kept fixed and another incoherent process leads the system back to its initial state $\hrh(2\tau_1+\tau_2+\tau_4)=\hrh_A$. The variation of mean energy of the system associated with each stroke is then given by
\begin{align}
W_{1} & =\bar{H}_{B}-\bar{H}_{A},\ \ \  Q_{2}=\bar{H}_{C}-\bar{H}_{B},\ \ \  
W_{3}=\bar{H}_{D}-\bar{H}_{C},\ \ \ 
Q_{4}=\bar{H}_{A}-\bar{H}_{D},
\label{eq:Q4}
\end{align}
where $\bar{H}_i=\Tr[\hH_i \hrh_i]$, $i=\{A,B,C,D\}$, denotes the mean energy at each point of the cycle, with $\hH_A=\hH_D=\hH(0)$ and $\hH_B=\hH_C=\hH(\tau_1)$. By this definition, useful energy is extracted from the system if the total work fulfills $W=W_1+W_3<0$. Consequently, the efficiency of the QOC in the heat engine configuration is given by the ratio $\mathcal{E}=-W/Q_{\textrm{abs}}$, where $Q_{\textrm{abs}}>0$ is the total energy that is incoherenlty absorbed by the system during the cycle. It is important to remark that the above development, which will guide our investigations hereafter, is simplified and does not take into account possible energy costs associated to the generation of the non-unitary processes through collisional models~\cite{Chiara2018}.

In a standard Otto cycle, the underlying non-unitary dynamics is provided by letting the work substance interact with thermal reservoirs, so that $\hrh_A$ and $\hrh_C$ are Gibbs states in the corresponding instantaneous eigenbasis, $\{\ket{E_{j}(0)}\}$ and $\{\ket{E_{j}(\tau_1)}\}$, respectively. In this case, the efficiency of the cycle is limited by the Carnot efficiency, $\mathcal{E}_{\textrm{Otto}}\leq \mathcal{E}_{\textrm{Carnot}}=1-T_{\textrm{c}}/T_{\textrm{h}}$, where $T_{\textrm{c/h}}$ is the temperature of the cold/hot reservoir. For a quantum mechanical system, such a bound may be violated upon a combination of non-adiabatic unitary dynamics and the presence of carefully engineered coherences of $\hrh_A$ or $\hrh_C$ as it will become clear later \footnote{Alternatively, effective negative temperatures may also contribute to an enhancement of the heat engine efficiency in the non-adiabatic regime for finite-dimensional quantum systems, see e.g. Ref.~\cite{Assis2019}.}. We expand $\hrh_A$ and $\hrh_C$ in the initial eigenbasis as 
\begin{align}
\hrh_{A}&=\sum_{j,k}p_{jk}\kb{E_{j}(0)}{E_{k}(0)},\\
\hrh_{C}&=\sum_{j,k}q_{jk}\kb{E_{j}(0)}{E_{k}(0)},
\label{eq:rhoC}
\end{align}
with coefficients $\{p_{jk}\}$ and $\{q_{jk}\}$, respectively. Note that the coefficients $\{q'_{jk}\}$ of $\hrh_C$ in the time-evolved eigenbasis $\{\ket{E_{j}(\tau_1)}\}$ are related to $\{q_{jk}\}$ as $q'_{jk}=\sum_{n,m}q_{nm}\bk{E_{j}(\tau_1)}{E_{n}(0)}\bk{E_{m}(0)}{E_{k}(\tau_1)}$.

Hereafter, we focus on the limiting case of non-adiabatic dynamics defined by the quench regime of strokes $1$ and $3$, so that one can perform the simplification
\begin{equation}
\hU(\tau_1)\ket{E_{j}(0)}\approx e^{-i\theta_{j}(\tau_1)}\ket{E_{j}(0)},\label{eq:quench}
\end{equation}
with $\theta_{j}(\tau_1)$ being a global dynamical phase. Namely, the changes in $\hH(t)$ are performed much faster than the response time of the system. In opposition to the quasi-static (adiabatic) limit, $\hU(\tau_1)\ket{E_{j}(0)}\propto \ket{E_{j}(\tau_1)}$, the mean energies $\bar{H}_B$ and $\bar{H}_D$ in the quench regime acquire a dependence on the coherences $\{p_{jk}\}$ and $\{q'_{jk}\}$. Using the correspondence between $\{q'_{jk}\}$ and $\{q_{jk}\}$ defined above, the set of equations~\eqref{eq:Q4} becomes
\begin{align}
W_{1} & =\sum_{j,k}p_{jk}e^{-i[\theta_{j}(\tau_1)-\theta_{k}(\tau_1)]}H_{kj}^{(0)}(\tau_1)-\sum_{j}p_{jj}E_{j}(0),\nonumber\\
Q_{2} & =\sum_{j,k}\left[q_{jk}-p_{jk}e^{-i[\theta_{j}(\tau_1)-\theta_{k}(\tau)]}\right]H_{kj}^{(0)}(\tau_1), \nonumber\\
W_{3} & =\sum_{j}q_{jj}E_{j}(0)-\sum_{j,k}q_{jk}H_{kj}^{(0)}(\tau_1),\nonumber\\
Q_{4} & =\sum_{j}(p_{jj}-q_{jj})E_{j}(0),\label{eq:Q4a}
\end{align}
where we defined the matrix elements $H_{kj}^{(0)}(\tau_1)=\mel{E_{k}(0)}{\hH(\tau_1)}{E_{j}(0)}$, the eigenvalues $E_{j}(t)=\mel{E_{j}(t)}{\hH(t)}{E_{j}(t)}$, and used the expansion of $\hrh_C$ in the initial eigenbasis.

With the vibrational mode as a work substance, we consider the general case where the unitary processes of the QOC are implemented by frequency modulation and displacement of the trap. The Hamiltonian may be written in a compact form as
\begin{equation}
\hH(t)=\hbar\nu(t)\left[\ha_{\textrm{d}}^{\dagger}(t)\ha_{\textrm{d}}(t)+\frac{1}{2}\right],\label{eq:Hfmd3}
\end{equation}
where $\nu(t)$ is the time-dependent angular frequency of the trap assuming the values $\nu_0\equiv\nu(0)$ and $\nu_1\equiv\nu(\tau_1)$. Also in equation~\eqref{eq:Hfmd3}, $\ha_{\textrm{d}}(t)$ and $\ha_{\textrm{d}}^{\dagger}(t)$ are displaced squeezed modes arising from the dynamical changes of the trap potential, such that by writing $\ha_0\equiv\ha_{\textrm{d}}(0)$ and $\hdgg{a}_0\equiv \ha_{\textrm{d}}^{\dagger}(0)$, the time-evolved modes at the instant $\tau_1$ become
\begin{align}
\ha_1 \equiv \ha_{\textrm{d}}(\tau_1) & =\cosh[r(\tau_1)]\ha_{0}+\sinh[r(\tau_1)]\ha_{0}^{\dagger}+\frac{\zeta_1}{\nu_1},\nonumber \\
\ha^{\dagger}_1 \equiv \ha_{\textrm{d}}^{\dagger}(\tau_1) & =\cosh[r(\tau_1)]\ha_{0}^{\dagger}+\sinh[r(\tau_1)]\ha_{0}+\frac{\zeta^{*}_1}{\nu_1},\label{eq:displadder}
\end{align}
where $r(\tau_1)=\ln(\nu_1/\nu_0)/2$ is a real-valued squeezing parameter and $\zeta_1/\nu_1$ is a complex-valued displacement that is introduced to the initial modes during stroke $1$. Thus, note that such time-evolved modes are produced by the application of the operator $\hdgg{S_{0}}[r(\tau_1)]\hdgg D_{0}(\zeta_1/\nu_1)$ on the initial modes. 
For ease of notation, we have omitted the internal degrees of freedom of the ion in equation~\eqref{eq:Hfmd3} since they are only accessed during the non-unitary strokes of the QOC. 

According to equation~\eqref{eq:Hfmd3}, the initial eigenbasis of the vibrational mode of the ion is formed by the number states $\ket{E_{j}(0)}=\ket{j_{\text{v}}}$, with corresponding eigenvalues $E_{j}(0)=\hbar\nu_{0}(j+1/2)$. The time-evolved eigenbasis at $t=\tau_1$ is composed by the displaced and squeezed number states
$\{\ket{E_{j}(\tau_1)} =\hdgg S_{0}[r(\tau_1)]\hdgg D_{0}(\zeta_1/\nu_1)\ket{j_{\text{v}}}\}$, with eigenvalues $\{E_{j}(\tau_1)=\hbar\nu_1(j+1/2)\}$. Consequently, it follows from equations~\eqref{eq:Q4a} that the total work and incoherent energies that are exchanged in the quench regime of the QOC read
\begin{align}
W & =-\hbar\left\{ \nu_{0}-(1-\chi)\nu_{1}\cosh[2r(\tau_1)]\right\} \left(\bar{n}_{A}-\bar{n}_{C}\right),\label{eq:WExb2}\\
Q_{2} & =-\hbar(1-\chi)\nu_{1}\cosh[2r(\tau_1)]\left(\bar{n}_{A}-\bar{n}_{C}\right),\label{eq:Q2Exb2}\\
Q_{4} & =\hbar\nu_{0}\left(\bar{n}_{A}-\bar{n}_{C}\right),\label{eq:Q4Exb2}
\end{align}
where we defined the average occupation number of states $\hrh_A$ and $\hrh_C$ in the initial eigenbasis as $\bar{n}_A=\sum_{j}p_{jj}j\kb{j_{\text{v}}}{j_{\text{v}}}$ and $\bar{n}_C=\sum_{j}q_{jj}j\kb{j_{\text{v}}}{j_{\text{v}}}$, respectively. Also, we introduced the dimensionless parameter
\begin{equation}
\chi=\frac{\chi_{C}^{(1)}+\chi_{C}^{(2)}-\chi_{A}^{(1)}-\chi_{A}^{(2)}}{\nu_{1}\cosh[2r(\tau_1)]\left(\bar{n}_{A}-\bar{n}_{C}\right)},\label{eq:shiftchi}
\end{equation}
which in turn depends on the angular frequencies
\begin{align}
\chi_{A}^{(1)} & =2\nu_{1}\textrm{Re}\left[\kappa_{1}\sum_{j=0}^{\infty}p_{jj+1}e^{i[\theta_{j+1}(\tau_1)-\theta_{j}(\tau_1)]}\sqrt{j+1}\right],\label{eq:chi1A}\\
\chi_{A}^{(2)} & =\nu_{1}\sinh[2r(\tau_1)]\textrm{Re\ensuremath{\left[\sum_{j=0}^{\infty}p_{jj+2}e^{i[\theta_{j+2}(\tau_1)-\theta_{j}(\tau_1)]}\sqrt{(j+1)(j+2)}\right]}},\label{eq:chi2A}\\
\chi_{C}^{(1)}&=2\nu_{1}\textrm{Re}\left[\kappa_{1}\sum_{j=0}^{\infty}q_{jj+1}\sqrt{j+1}\right],\label{eq:chi1C}\\
\chi_{C}^{(2)}&=\nu_{1}\sinh[2r(\tau_1)]\textrm{Re\ensuremath{\left[\sum_{j=0}^{\infty}q_{jj+2}\sqrt{(j+1)(j+2)}\right]},}\label{eq:chi2C}
\end{align}
with $\kappa_1\equiv\frac{\zeta_1}{\nu_1}\cosh[r(\tau_1)]+\frac{\zeta^{*}_1}{\nu_1}\sinh[r(\tau_1)]$. Therefore, for a fixed difference $\bar{n}_{A}-\bar{n}_{C}$, the parameter $\chi$ effectively rescales the trap frequency $\nu_{1}$ according to the displacement, squeezing, and the residual coherences of the dissipatively generated states in the QOC. More precisely, $\chi_{A/C}^{(1)}$ account for contributions from the displacement of the trap and the first neighbour coherences, while $\chi_{A/C}^{(2)}$ are produced by the frequency modulation of the trap and second neighbour coherences.

The conditions considered so far promote a diverse scenario for incoherently absorbed energy and work extraction in the quench regime of the QOC, being assisted by the control of $\chi$ through reservoir engineering. From equations~\eqref{eq:WExb2}--\eqref{eq:Q4Exb2} and by imposing $W<0$, we can write the efficiency of the QOC in the quench regime enclosing all cases as
\begin{align} \mathcal{E} & =\begin{cases} 1-\frac{\nu_{1}\textrm{cosh}[2r(\tau_1)](1-\chi)}{\nu_{0}}, & \bar{n}_{A}>\bar{n}_{C}\textrm{\textrm{ and} }1-\frac{\nu_{0}}{\nu_{1}}\textrm{sech}[2r(\tau_1)]<\chi<1,\\ 1, & \bar{n}_{A}>\bar{n}_{C}\textrm{\textrm{ and} }\chi\geq1,\\ 1-\frac{\nu_{0}\textrm{sech}[2r(\tau_1)]}{\nu_{1}(1-\chi)}, & \bar{n}_{A}<\bar{n}_{C}\textrm{ \textrm{and} }\chi<1-\frac{\nu_{0}}{\nu_{1}}\textrm{sech}[2r(\tau_1)],\\ 0, & \chi=1-\frac{\nu_{0}}{\nu_{1}}\textrm{sech}[2r(\tau_1)]. \end{cases}\label{eq:effquenchall} 
\end{align}
For reference, we introduce the efficiency of a standard QOC with a trapped ion under frequency modulation in the quasi-static regime as~\cite{Abah2012,Kosloff2017}
\begin{equation}
\mathcal{E}_{\textrm{Otto}}=1-\left[\frac{\nu_{1}}{\nu_{0}}\Theta(\nu_0-\nu_1)+\frac{\nu_{0}}{\nu_{1}}\Theta(\nu_1-\nu_0)\right],\label{eq:effOtto}
\end{equation}
with $\Theta(x)$ being a step function. When complete thermalization processes dictate the state production in strokes $2$ and $4$, the efficiency $\mathcal{E}_{\textrm{Otto}}$ is the maximum allowed value.

\begin{figure}[ht]
\begin{center}
\includegraphics[width=\linewidth]{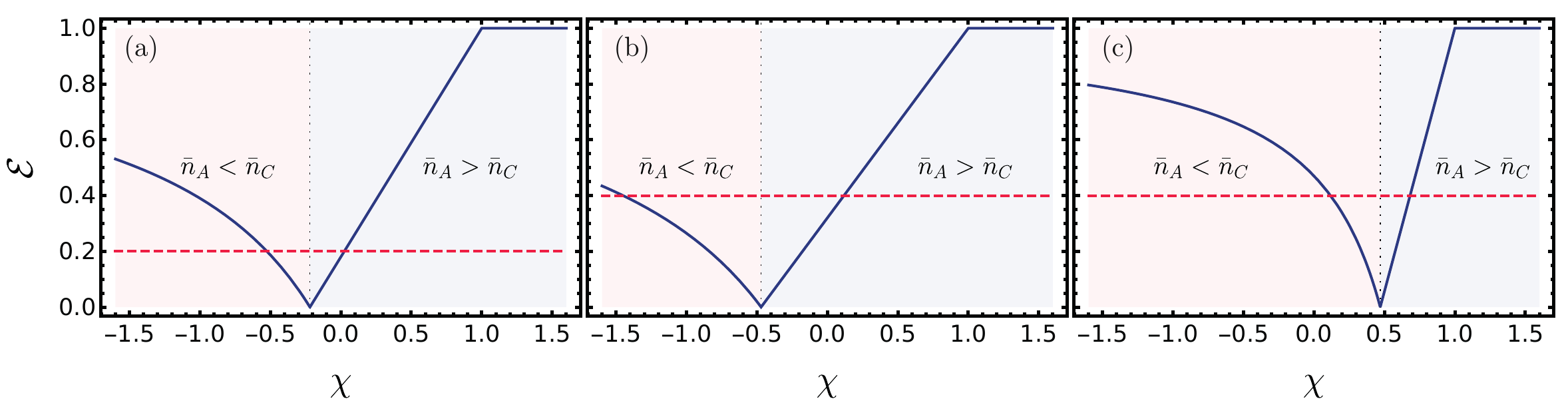} 
\subfloat{\label{fig:effquencha}}
\subfloat{\label{fig:effquenchb}}
\subfloat{\label{fig:effquenchc}}
\caption{Efficiency $\mathcal{E}$ of the QOC (in the quench regime), given by equation~\eqref{eq:effquenchall}, as function of $\chi$ for different values of $\nu_{1}/\nu_{0}$. The horizontal dashed lines correspond to the efficiency $\mathcal{E}_{\textrm{Otto}}$ of a standard QOC, given by equation~\eqref{eq:effOtto}. The vertical dotted lines delimit the work extraction regions when $\bar{n}_{A}<\bar{n}_{C}$ and $\bar{n}_{A}>\bar{n}_{C}$. In the plots we have chosen (a) $\nu_{1}/\nu_{0}=0.800$,
(b) $\nu_{1}/\nu_{0}=0.600$, and (c) $\nu_{1}/\nu_{0}=1.67$, providing better efficiencies than $\mathcal{E}_{\textrm{Otto}}$ if:
$\chi>2.44\times 10^{-2}$ or $\chi<-5.24\times 10^{-1}$ [case (a)]; $\chi>1.18 \times 10^{-1}$
or $\chi<-1.45$ [case (b)]; $\chi>6.82\times 10^{-1}$ or $\chi<1.18\times 10^{-1}$ [case (c)].
}
\label{fig:effquench} 
\end{center}
\end{figure}

In Figure~\ref{fig:effquench}, we used equation~\eqref{eq:effquenchall} to show the behavior of $\mathcal{E}$ as function of $\chi$. This allowed us to show the conditions where the efficiency surpasses $\mathcal{E}_{\textrm{Otto}}$ for different values of $\nu_{1}/\nu_{0}$. In figures~\ref{fig:effquencha} and~\ref{fig:effquenchb}, where $\nu_{1}/\nu_{0}<1$, it turns out that for $\bar{n}_{A}>\bar{n}_{C}$, the efficiency surpasses $\mathcal{E}_{\textrm{Otto}}$  as long as $\chi>1-\textrm{sech}[2r(\tau_1)]$. In particular, it reaches unity if $\chi\geq1$. On the other hand, for $\bar{n}_{A}<\bar{n}_{C}$, surpassing efficiencies are obtained if $\chi<1-\left(\nu_{0}/\nu_{1}\right)^{2}\textrm{sech}[2r(\tau_1)]$. When $\chi=1-\left(\nu_{0}/\nu_{1}\right)\textrm{sech}[2r(\tau_1)]$, a null efficiency is generated since this is the limiting case where $W=0$. Figure~\ref{fig:effquenchc} shows a situation where $\nu_{1}/\nu_{0}>1$. In this case, when $\bar{n}_{A}>\bar{n}_{C}$, one obtains $\mathcal{E}>\mathcal{E}_{\textrm{Otto}}$ for values of $\chi>1-\left(\nu_{0}/\nu_{1}\right)^{2}\textrm{sech}[2r(\tau_1)]$, also reaching unity when $\chi\geq1$. For $\bar{n}_{A}<\bar{n}_{C}$, $\mathcal{E}>\mathcal{E}_{\textrm{Otto}}$ is achieved provided $\chi<1-\textrm{sech}[2r(\tau_1)]$.

\subsection{Non-unitary state generation\label{sec:nonunit}}

As in section~\ref{sec:appl}, we  consider the case of a $4$--level trapped ion. Here, we choose a thermal state at the beginning of the cycle, $\hrh_A$, and a squeezed coherent state produced during stroke $2$, $\hrh_C$. Their synthesis was discussed in section~\ref{sec:appl} particularly in figure~\ref{fig:fidNotto}. However, it is worth mentioning that the squeezed coherent states are now generated in the time-evolved basis $\{\ket{E_{j}(\tau_1)} =\hdgg S_{0}[r(\tau_1)]\hdgg D_{0}(\zeta_1/\nu_1)\ket{j_{\text{v}}}\}$. Consequently, it is possible to choose the right squeezing parameters to make this state equivalent to coherent states $\hD_0(\alpha)\ket{0_{\textrm{v}}}$ in the initial energy eigenbasis $\{\ket{E_j(0)}=\ket{j_{\textrm{v}}}\}$. 
Therefore, we can write the states $\hrh_A$ and $\hrh_C$ as
\begin{align}
\hrh_{A}&=\sum_{j=0}^{\infty}p_{jj}\kb{j_{\text{v}}}{j_{\text{v}}},\ \ \ \ \ \ \ \ \ \ \  p_{jj}=\frac{\bar{n}_{A}^{j}}{(\bar{n}_{A}+1)^{j+1}},\label{eq:pjjtherm}\\ 
\hrh_{C}&=\sum_{j,k=0}^{\infty}q_{jk}\kb{j_{\text{v}}}{k_{\text{v}}},\ \ \ \ \ \ \ \ \ q_{jk}=e^{-\n{\alpha}^{2}}\frac{\alpha^{j}\alpha^{*k}}{\sqrt{j!k!}},\label{eq:rhoCcoh}
\end{align}
with average occupation number $\bar{n}_A$ and $\bar{n}_C=\n{\alpha}^2$ in the initial eigenbasis. While this choice of $\hrh_A$ yields $\chi_{A}^{(1)}=\chi_{A}^{(2)}=0$ in equations~\eqref{eq:chi1A} and~\eqref{eq:chi2A}, the choice of $\hrh_C$ yields  
\begin{align} 
\chi_{C}^{(1)} & =2\nu_{1}\textrm{Re}\left(\kappa_{1}\alpha^{*}\right),\label{eq:chi1Cex}\\ \chi_{C}^{(2)} & =\nu_{1}\sinh[2r(\tau_1)]\textrm{Re\ensuremath{\left(\alpha^{*2}\right)},}\label{eq:chi2Cex} \end{align}
so that the parameter $\chi$ becomes 
\begin{equation} \chi=\frac{2\text{sech}[2r(\tau_1)]\textrm{Re}\left(\kappa_{1}\alpha^{*}\right)+\text{\ensuremath{\tanh}}[2r(\tau_1)]\textrm{Re\ensuremath{\left(\alpha^{*2}\right)}}}{\left(\bar{n}_{A}-\n{\alpha}^{2}\right)}.\label{eq:shiftchiex} 
\end{equation}
Notice that the displacement $\zeta_1/\nu_1\neq0$ does not influence the values of $\chi$ if $\textrm{Re}\left(\kappa_{1}\alpha^{*}\right)=0$, which occurs if the phase difference between $\zeta_1/\nu_1$ and $\alpha$ is an integer multiple of $\pi$. 

\begin{figure}[ht]
\begin{center}
\includegraphics[width=0.5\linewidth]{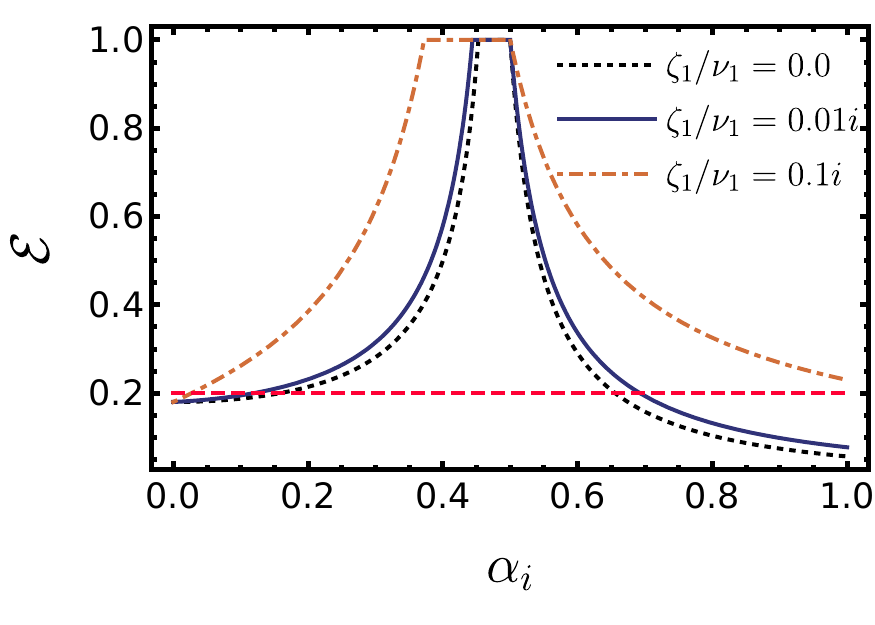} 
\caption{Quench regime efficiency $\mathcal{E}$ as function of $\alpha_i$ for different values of $\zeta_1/\nu_1$. The other parameters are chosen as $\nu_1/\nu_0=0.80$ and $\bar{n}_A=0.25$. The horizontal dashed lines gives the corresponding standard Otto efficiencies $\mathcal{E}_{\text{Otto}}$.}
\label{fig:effalpha} 
\end{center}
\end{figure}

Figure~\ref{fig:effalpha} shows the quench efficiency $\mathcal{E}$ as a function of $\alpha_i$ for different values of $\zeta_1/\nu_1$ and $\alpha=i\alpha_i$. We can observe that $\mathcal{E}_{\textrm{Otto}}$ is surpassed for a broad range of values of $\alpha_i$, even without displacement, $\zeta_1/\nu_1=0$. This is attributed solely to the term $\text{\ensuremath{\tanh}}[2r(\tau_1)]\textrm{Re\ensuremath{\left(\alpha^{*2}\right)}}$ in equation~\eqref{eq:shiftchiex}. However, we note that the regions where $\mathcal{E}>\mathcal{E}_{\textrm{Otto}}$ may be extended with $\zeta_1/\nu_1\neq0$. For the chosen parameters, therefore, displacements of the trap promote the amplification of quantum effects that are observed in the heat engine.

It is important to remark that $\mathcal{E}=1$ has already been discussed in the scope of non-thermal baths  \cite{Niedenzu2016,Niedenzu2018}, and our protocol provides a recipe to observe it in the context of trapped ions, which is a promising setup to test fundamentals of quantum thermodynamics \cite{Huber2008,Abah2012,Cifuentes2016}. We also emphasize that sequential interactions are at the core of the so-called micromaser~\cite{Scully1967}. This setup is central to quantum optics and much of what is currently known about the role of coherences and photon statistics comes from investigations on micromasers. Typically, this setup consists of the passage of single atoms through a cavity sustaining modes of the electromagnetic field~\cite{Scully1967}. The latter are described as quantized bosonic systems just like the vibrational mode of a trapped ion. In this way, much of what we do here has a formal relation with micromasers. However, in the trapped ion setup, the simple choice of the laser frequency can lead to a great variety of effective interactions which do not easily appear in the micromaser, unless additional fields and more general energy level structures are supplied. One interesting route for future investigation would be the generalization of the multibath engineering scheme presented here to clusters formed by different numbers of ions in the trap. This takes direct inspiration from the clustering of atoms in the micromaser as proposed in~\cite{Dag2016}. In this work, it is shown that different clusterings represent fuels of different performances for the design of thermal and nonthermal machines. Coherence is once again a key ingredient in the state space of more than one atom, i.e., the cluster.

\section{Conclusions} \label{sec:conc}

In conclusion, we have presented a collisional protocol for multireservoir engineering to generate the vibrational state of a trapped ion. The method makes use of a $d$--level trapped ion interacting with pulsed lasers in the resolved sideband regime. One of the main results that follow from our approach is the unique possibility of engineering thermal environments with controlled temperatures. We then proposed the use of asymptotic states produced by engineered multireservoirs to surpass the thermal efficiency of Otto cycles in quantum thermodynamics. This complements a few previous proposals in the subject~\cite{Assis2019,Niedenzu2016,Niedenzu2018}. In general, the multireservoir setup can be used in a large breadth of applications relying on the use of state synthesis, quantum simulation of open systems as well as scenarios where multiple reservoirs are needed such as in the quantum transport~\cite{Moreira2021,RomanAncheyta2021}. We hope our work can motivate further applications of reservoir-induced dynamics, in the scope, for instance, of quantum computation driven by dissipation~\cite{Verstraete2009} and non equilibrium transport resulting from multiple currents~\cite{Nicacio2015,Xuereb2015,Falasco2015}.
\section*{Acknowledgements} \label{sec:acknowledgements}
W.S.T. acknowledges Fundação de Amparo à Pesquisa do Estado de São Paulo (FAPESP) for financial support through Grant No. 2017/09058-2. F.L.S. acknowledges partial support from the Brazilian National Institute of Science and Technology of Quantum Information (CNPq INCT-IQ 465469/2014-0), CNPq (Grant No. 305723/2020-0), CAPES/PrInt (88881.310346/2018-01). This work was supported by the UK EPSRC Hub in Quantum Computing and Simulation (EP/T001062/1).

\footnotesize
\bibliography{refs2}

\end{document}